\documentstyle[a4,12pt]{article}

\newcommand{\nit}{\noindent}

\newcommand{\np}{\newpage}
\newcommand{\dsp}{\displaystyle}
\newcommand{\vs}[1]{\vspace{#1 ex}}
\newcommand{\hs}[1]{\hspace{#1 em}}
\newcommand{\bfr}{\begin{flushright}}
\newcommand{\efr}{\end{flushright}}
\newcommand{\bc}{\begin{center}}
\newcommand{\ec}{\end{center}}
\newcommand{\ben}{\begin{enumerate}}
\newcommand{\een}{\end{enumerate}}

\newcommand{\be}{\begin{equation}}
\newcommand{\ee}{\end{equation}}
\newcommand{\ba}{\begin{array}}
\newcommand{\ea}{\end{array}}
\newcommand{\ct}{\cite}
\newcommand{\bit}{\bibitem}
\newcommand{\dd}[2]{\frac{\partial{#1}}{\partial{#2}}}
\newcommand{\ag}{\alpha}

\newcommand{\del}{\delta}
\newcommand{\eps}{\epsilon}
\newcommand{\ve}{\varepsilon}

\newcommand{\thg}{\theta}
\newcommand{\kg}{\kappa}
\newcommand{\lb}{\lambda}
\newcommand{\sg}{\sigma}
\newcommand{\rg}{\rho}

\newcommand{\fg}{\phi}

\newcommand{\Del}{\Delta}

\newcommand{\bz}{\bar{z}}

\newcommand{\lh}{\left(}
\newcommand{\rh}{\right)}
\newcommand{\ld}{\left.}
\newcommand{\rd}{\right.}

\newcommand{\lrder}{\stackrel{\leftrightarrow}{\der}}

\newcommand{\cL}{{\cal L}}

\newcommand{\der}{\partial}

\begin{document}

\pagestyle{empty}

\begin{flushright}
NORDITA 2005-8
\end{flushright}
\vs{2}

\begin{center}
{\Large {\bf Conformal fluid dynamics}}\\
\vs{7}

{\bf P.D.\ Jarvis}$^*$ \\
\vs{2} 
University of Tasmania \\
\vs{1} 
Hobart, Tasmania \\
\vs{2}

and \\
\vs{2}
 
{\bf J.W.\ van Holten}$^{**\, \dagger}$ \\
\vs{2}
Nordita \\
\vs{1}
Copenhagen DK\\
\vs{3}

January 27, 2005
\end{center}
\vs{3}

\nit
{\footnotesize{ {\bf Abstract} \\
We present a conformal theory of a dissipationless relativistic fluid in 2
space-time dimensions. The theory carries with it a representation of the
algebra of 2-$D$ area-preserving diffeomorphisms in the target space of
the complex scalar potentials. A complete canonical description is given,
and the central charge of the current algebra is calculated. The passage
to the quantum theory is discussed in some detail; as a result of
operator ordering problems, full quantization at the level of the fields
is as yet an open problem.
}}

\vfill
\footnoterule

\nit
{\footnotesize{ $^*$ {\tt e-mail: peter.jarvis@utas.edu.au } \\
                      $^{**}$ {\tt e-mail: v.holten@nikhef.nl} \\
                      $^{\dagger}$ on leave from NIKHEF, Amsterdam NL }}

\np

\pagestyle{plain}
\pagenumbering{arabic}

\section{Fluid dynamics as a lagrangean field theory \label{s.1}}

Non-dissipative fluid mechanics in either a relativistic or non-relativistic framework
can be formulated as a lagrangean field theory \ct{jackiw,jnpp}. In a relativistic
context, the relevant physical degrees of freedom are described by the time-like
four-current density $j^{\mu}(x)$, related to the scalar density $\rg(x)$ by
\be
g_{\mu\nu}\, j^{\mu} j^{\nu} = - \rg^2.
\label{1.1}
\ee
The velocity field $u^{\mu}(x)$ then is related to the current density by
\be
j^{\mu} = \rg u^{\mu}, \hs{2}  g_{\mu\nu}\, u^{\mu} u^{\nu} = -1.
\label{1.2}
\ee
The essential physical aspects of fluid dynamics in Minkowski space are the
conservation laws as represented by vanishing divergence of the fluid current
and the energy-momentum tensor:
\be
\der_{\mu}  j^{\mu} = 0, \hs{2} \der_{\mu} T^{\mu\nu} = 0.
\label{1.3}
\ee
In the non-relativistic limit these reduce to the Bernouilli and Euler equations.
In addition, there is an equation of state relating the pressure $p$ and energy
density $\ve = f(\rg)$. It is straightforward to  generalize this description to a
general relativistic context so as to include the gravitational field \ct{weinberg}.
A lagrangean formulation of a fully relativistic theory, including the gravitational
field, in $n$ space-time dimensions is given by the action
\be
S = \int d^nx\, \sqrt{-g} \lh -\frac{1}{8 \pi G}\, R + \cL_{fluid} \rh,
\label{1.4}
\ee
where the first term is the Einstein-action for general relativity, and the lagangean
density for the fluid is
\be
\cL_{fluid} = - j \cdot \lh \der \thg + i \bz \der z - i z \der \bz \rh - f(\rg).
\label{1.5}
\ee
Here $(\thg, \bz, z)$ are auxiliary fields acting as langrange multipliers imposing
the correct physical conditions on the current density. Strictly speaking, the
action (\ref{1.5}) is motivated from fluid dynamics in $n =4$, but it can consistently
be continued to other values of $n$ as well.

The field equations derived from the action (\ref{1.4}) by varying the current
components $j^{\mu}$ and the potentials $(\thg, \bz, z)$ lead to an equation
for the current
\be
\frac{f^{\prime}(\rg)}{\rg}\, j_{\mu} = \der_{\mu} \thg
 + i \bz \der_{\mu} z - i z \der_{\mu} \bz,
\label{1.6}
\ee
subject to the conditions
\be
D \cdot j = 0, \hs{1} j \cdot \der z = j \cdot \der \bz = 0.
\label{1.7}
\ee
These are supplemented by the Einstein equations
\be
R_{\mu\nu} - \frac{1}{2}\, g_{\mu\nu}\, R = - 8\pi G\, T_{\mu\nu},
\label{1.8}
\ee
with the energy-momentum tensor taking the perfect-fluid form
\be
T_{\mu\nu} = p g_{\mu\nu} + (\ve + p) u_{\mu} u_{\nu}.
\label{1.9}
\ee
In this expression the energy density and pressure are given by
\be
\ve = f(\rg), \hs{2} p = \rg f^{\prime}(\rg) - f(\rg).
\label{1.10}
\ee
A typical equation of state is found by taking a power law for the energy density
\be
f(\rg) = \ag \rg^{1 + \eta} \hs{1} \Rightarrow \hs{1} p = \eta\, \ve.
\label{1.11}
\ee
This is the type of equation of state familiar from applications in cosmology and
astrophysics.

Eqs.\ (\ref{1.7}) for the potentials $(\bz, z)$ state that these auxiliary fields are
constant in a comoving frame:
\be
u \cdot \der z = 0 \hs{1} \Leftrightarrow \hs{1}
\frac{dz}{dt} = \dd{z}{t} + {\bf v} \cdot \nabla z = 0,
\label{1.12}
\ee
and similarly for the complex conjugate potential. As a result, there is an infinite
set of covariantly conserved currents of the form \ct{nyawetal}
\be
J[G] = 2 G(\bz, z)  j_{\mu}, \hs{2} D \cdot J[G] = 0.
\label{1.13}
\ee
The factor $2$ results from Noether's theorem, applied to the invariance of
the action under the infinitesimal transformations
\be
\del_G \thg = 2 G - z G_{,z} - \bz G_{, \bz}, \hs{1}
\del_G z = - i G_{,\bz}, \hs{1} \del_G \bz = i G_{,z},
\label{1.14}
\ee
whilst $\del_G j^{\mu} = \del_G g_{\mu\nu} = 0$. The commutator algebra of
these infinitesimal transformations is closed, with the composition rule
\be
\left[ \del_G, \del_{G^{\prime}} \right] = \del_{G^{\prime\prime}}, \hs{2}
G^{\prime\prime} = i \lh \dd{G}{z} \dd{G^{\prime}}{\bz} - \dd{G}{\bz} \dd{G^{\prime}}{z} \rh.
\label{1.14.1}
\ee
It is readily checked that the transformations (\ref{1.14}) represent the invariances
of the one-form
\be
J = d \thg + i \bz dz - i z d \bz.
\label{1.15}
\ee
This also implies the existence of an invariant two-form
\be
A = \frac{1}{2}\, d \wedge J = i d\bz \wedge dz,
\label{1.16}
\ee
representing the area in the space of complex potentials. Area-preserving
diffeomorphisms have been studied in a different context as a symmetry
of the base-space manifold of the relativistic membrane in ref.\
\ct{hoppe,dewit-nicolai}. In contrast, here the diffeomorphisms are realized
directly on the dynamical variables. It is an interesting question what happens
to these transformations in a quantum theory, where these variables become
operators.

\section{Conformal fluid dynamics \label{s.2}}

For specific equations of state, the classical relativistic fluid models (\ref{1.5})
are conformally invariant. Indeed, it is easily established that the trace of the
energy-momentum tensor (\ref{1.9}) vanishes for fluids in $n$ space-time
dimensions for energy densities
\be
\ve = f(\rg) = \ag \rg^{\frac{n}{n-1}} \hs{1} \Rightarrow \hs{1} T_{\mu}^{\;\;\mu} = 0.
\label{2.1}
\ee
For such fluids the equation of state parameter is
\be
\eta  = \frac{1}{n-1}.
\label{2.2}
\ee
In 4 space-time this implies
\be
\eta = \frac{1}{3} \hs{1} \Rightarrow \hs{1} p = \frac{1}{3}\, \ve
\hs{3} (n = 4).
\label{2.3}
\ee
This is the indeed the equation of state for a gas of free massless particles, like
photons or massless neutrinos.

In two space-time dimensions a conformal fluid model has a somewhat
unusual equation of state with $\eta = 1$:
\be
p = \ve \hs{3} (n = 2).
\label{2.4}
\ee
The action in this case reads
\be
\ba{lll}
S & = & \dsp{ \int d^2 x\, \sqrt{-g} \left[- j^{\mu} \lh \der_{\mu} \thg
 + i \bz \der_{\mu} z - i z \der_{\mu} \bz \rh + \ag g_{\mu\nu} j^{\mu} j^{\nu} \right] }\\
 & & \\
 & \simeq &  \dsp{ - \frac{1}{4\ag}\, \int d^2 x\, \sqrt{-g} g^{\mu\nu}
 \lh \der_{\mu} \thg + i \bz \der_{\mu} z - i z \der_{\mu} \bz \rh
 \lh \der_{\nu} \thg + i \bz \der_{\nu} z - i z \der_{\nu} \bz \rh,  }
\ea
\label{2.4.1}
\ee
where the last line is obtained by algebraic elimination of the independent
current vector field $j^{\mu}$ by
\be
j_{\mu} = \frac{1}{2\ag} \lh \der_{\mu} \thg + i \bz \der_{\mu} z - i z \der_{\mu} \bz \rh.
\label{2.4.2}
\ee
If the normalization of the current is chosen such that $\ag = 1/2$,
one obtains standard kinetic terms for the scalar potential $\thg$.

A further important aspect of this 2-dimensional model is that one
can extend the model with a Wess-Zumino term in which the real
scalar $\thg$ couples to the invariant target-space area 2-form:
\be
\Del S = 2i \lb \int d^2 x\, \ve^{\mu\nu} \thg\, \der_{\mu} \bz \der_{\nu} z.
\label{2.5}
\ee
For $2 \ag \lb = \pm 1$ this implies that the current becomes self dual, or
anti-self dual, respectively. It is easily verified that the action (\ref{2.4.1})
and the Wess-Zumino term are invariant under local Weyl rescaling
\be
g_{\mu\nu}(x) \rightarrow g^{\prime}_{\mu\nu}(x) = e^{\sg(x)} g_{\mu\nu}(x),
\label{2.6}
\ee
keeping the scalars $(\thg, \bz, z)$ unchanged; this confirms the conformal
invariance of the model. Remarkably, the infinite set of transformations
(\ref{1.14}) also leaves the Wess-Zumino term invariant, up to a boundary
term. In part, this results from the invariance of the area two-form $A$;
the non-trivial part is, that the transformation $\del_G \thg$ by itself
produces a total derivative.

Finally we observe, that the energy-momentum tensor of this theory is not
changed by the addition of the Wess-Zumino term and with $\ag = 1/2$
takes the Sugawara form
\be
T_{\mu\nu} = j_{\mu} j_{\nu} - \frac{1}{2}\, g_{\mu\nu} g^{\kg \lb} j_{\kg} j_{\lb},
\label{2.7}
\ee
which is traceless in 2-dimensional space-time.

\section{Canonical light-cone formulation \label{s.3}}

In this section we present the canonical formulation of 2-dimensional
 conformal fluid dynamics in the light-cone formulation, using light-cone
co-ordinates and derivatives
\be
x^{\pm} = \frac{x^0 \pm x^1}{\sqrt{2}}, \hs{2}
\der_{\pm} = \frac{\der_0 \pm \der_1}{\sqrt{2}},
\label{3.1}
\ee
such that in the conformal gauge the space-like line element is
\be
ds^2 = - 2 e^{\fg(x)}\, dx^+ dx^-.
\label{3.1.1}
\ee
Because of the conformal invariance, the action is independent of the
2-dimensional conformal gravitational field component $\fg(x)$. Therefore,
using the canonical value $\ag = 1/2$, the full
action for the conformal theory takes the form
\be
\ba{lll}
S_{lc} & = & \dsp{
  \int dx^+ dx^- \left[ (\der_+ \thg + i \bz \der_+ z - i z \der_+ \bz)
 ( \der_- \thg + i \bz \der_- z - i z \der_- \bz) \rd }\\
 & & \\
 & & \dsp{ \ld \hs{5}
 +\, 2 i \lb \thg \lh \der_+ \bz \der_ - z - \der_- \bz \der_+ z \rh \right]. }
\ea
\label{3.2}
\ee
The field equations then include the self dual/anti-self dual current
conservation laws
\be
\ba{ll}
\der_- J_+ = 0, & \lb = +1; \\
 & \\
\der_+ J_- = 0, & \lb = -1,
\ea
\label{3.3}
\ee
where the current components $J_{\pm}$ represent the fixed expressions
\be
J_{\pm} = \der_{\pm} \thg + i \bz  \der{_\pm} z - i z \der_{\pm} \bz.
\label{3.3.1}
\ee
These equations lead to the result
\be
\ba{ll}
\der_+ J_- = 2i \lh  \der_+ \bz \der_- z - \der_- \bz \der_+ z \rh, & \lb = +1; \\
 & \\
\der_- J_+ = -2i \lh \der_+ \bz \der_- z - \der_- \bz \der_+ z \rh, & \lb  = -1.
\ea
\label{3.3.2}
\ee
The other field equations, stating that the flow is orthogonal to the gradient
of the complex scalars, read
\be
\ba{ll}
J_- \der_+ z = J_- \der_+ \bz = 0, & \lb = +1; \\
 & \\
J_+ \der_- z = J_+ \der_- \bz = 0, & \lb = -1.
\ea
\label{3.4}
\ee
Together eqs.\ (\ref{3.3})-(\ref{3.4}) are seen to imply full current conservation
\be
\der_- J_+ = \der_+ J_- = 0, \hs{2} \lb = \pm 1.
\label{3.4.0}
\ee
Observe, that changing the value of $\lb$ form $+1$ to $-1$ is equivalent to
interchanging the role of $x^+$ and $x^-$. In the following we therefore restrict
ourselves to $\lb = +1$, taking $x^+$ as the light-cone time.

Finally, the energy-momentum tensor is traceless, and has only two
non-vanishing components
\be
T_{++} = J_+ J_+, \hs{2} T_{--} = J_- J_-.
\label{3.5}
\ee
Their divergence vanishes:
\be
\der_- T_{++} = \der_+ T_{--} = 0.
\label{3.6}
\ee
With $x^+$ as the light-cone time co-ordinate,  the action takes  the form
\be
S_{lc} = \int dx^+ dx^-\, A_i[\fg]\, \der_+ \fg_i,
\label{3.7}
\ee
where the fields are denoted collectively by $\fg_i = (\thg, z, \bz)$, and
where the field-dependent coefficients $A_i[\fg]$ read
\be
\ba{lll}
A_{\thg} & = & J_ - = \der_- \thg + i \bz \der_- z - i z \der_- \bz, \\
 & & \\
A_z & = & i \bz J_- - 2 i \lb \thg \der_- \bz, \\
 & & \\
A_{\bz} & = & -i z J_- + 2 i \lb \thg \der_- z,
\ea
\label{3.8}
\ee
where in the following we use $\lb = + 1$. We follow the Faddeev-Jackiw
procedure \ct{faddeev-jackiw}, and write the variations of the action in the
form
\be
\frac{\del S_{lc}}{\del \fg_i(x^+, x^-)}
 = \int dy^-\, F_{ij}(x^+; x^-, y^-)\, \der_+ \fg_j(x^+, y^-),
\label{3.9}
\ee
with the symplectic tensor $F_{ij}$ defined by
\be
F_{ij}(x^+; x^-, y^-) = \frac{\del A_j(x^+,y^-)}{\del \fg_i(x^+,x^-)}
 - \frac{\del A_i(x^+,x^-)}{\del \fg_j(x^+,y^-)}.
\label{3.10}
\ee
A list of explicit expressions for the components of $F_{ij}$ is given in
appendix \ref{a.1}. Leaving understood that all expressions are to be
evaluated at equal light-cone time $x^+$, the equations of motion can
then be written in the form
\be
\der_+ \fg_i(x^-) = \int dy^-\, F^{-1}_{ij}(x^-, y^-)\, \frac{\del S}{\del \fg_j(y^-)}
 = \left\{ \fg_i(x^-), S \right\},
\label{3.11}
\ee
with the Poisson brackets defined by the inverse of the symplectic tensor:
\be
\left\{ \fg_i(x^-), \fg_j(y^-) \right\} = F^{-1}_{ij} (x^-, y^-).
\label{3.12}
\ee
Note, that this procedure is closely related to the Schwinger source method,
but without the necessity for any explicit introduction of source terms.
\vs{1}

\nit
To calculate the Poisson brackets we use the components of the
symplectic tensor in app.\ \ref{a.1}, and observe that it takes the form
\be
F_{ij}(x^-, y^-) = - 2 S_{ij}(x^-, y^-)\, \del^{\, \prime} ( x^- - y^- )
 - 2i A_{ij}(x^-)\, \del ( x^- - y^- ),
\label{3.15}
\ee
with
\be
S_{ij}(x^-, y^-) = \lh \ba{ccc} 1 & i \bz(y^-) & - i z(y^-) \\
                          i \bz(x^-) & - \bz(x^-) \bz(y^-) & \bz(x^-) z(y^-) \\
                        - i z(x^-) & \bz(y^-) z(x^-) & -z(x^-) z(y^-) \ea \rh,
\label{3.16}
\ee
and
\be
A_{ij}(x^-) = 2 J_-(x^-) \lh \ba{ccc} 0 & 0 & 0 \\
                                                 0 & 0 & 1 \\
                                                 0 & -1 & 0 \ea \rh.
\label{3.17}
\ee
Its inverse is
\be
\ba{lll}
\left\{ \fg_i(x^-), \fg_j(y^-) \right\} & = & F^{-1}_{ij}(x^-, y^-) \\
 & & \\
 & = & \dsp{ - \frac{1}{2}\, M_{ij}(x^-, y^-)\, \eps(x^- - y^-) - \frac{1}{2}\, N_{ij}(x^-)\,
 \del (x^- - y^-), }
\ea
\label{3.18}
\ee
where $\eps(x - y)$ is the anti-symmetric Heavyside step function, and
\be
M_{ij}(x, y) = \lh \ba{ccc} 1 & 0 & 0 \\
                                          0 & 0 & 0 \\
                                          0 & 0 & 0 \ea \rh, \hs{1}
N_{ij}(x) = \frac{1}{2J_-(x)} \lh \ba{ccc} 0 & z(x) & \bz(x) \\
                                                            -z(x) & 0 & i \\
                                                            -\bz(x) & -i & 0 \ea \rh.
\label{3.19}
\ee
We can now read off the Poisson brackets of the various fields; the non-zero
elementary brackets are the following
\be
\ba{ll}
\dsp{ \left\{ \thg(x^-), \thg(y^-) \right\} =  - \frac{1}{2}\, \eps (x^- - y^- ),  }&
\dsp{ \left\{ z(x^-), \bz(y^-) \right\} = - \frac{i}{4J_-}\, \del(x^- - y^-),  }\\
 & \\
\dsp{ \left\{ z(x^-), \thg(y^-)\right\} = \frac{z}{4J_-}\, \del (x^- - y^-), }&
\dsp{ \left\{ \bz(x^-), \thg(y^-) \right\} = \frac{\bz}{4J_-}\, \del (x^- - y^-). }\\
\ea
\label{3.20}
\ee
With the aid of these expressions, the bracket of the fields with the right-moving
current component takes the form
\be
\left\{ \fg_i(x^-), J_-(y^-) \right\} = \frac{1}{2J_-}\, \der_- \fg_i\, \del(x^- - y^-),
\label{3.21}
\ee
and as a result
\be
\left\{ \fg_i(x^-), T_{--}(y^-) \right\} = \left\{ \fg_i(x^-), J_-^2(y^-) \right\}
 = \der_- \fg_i\, \del(x^- - y^-),
\label{3.22}
\ee
as expected. Finally, the equal light-cone time brackets of the currents
and energy-momentum tensor
themselves read
\be
\ba{l}
\dsp{ \left\{ J_-(x^-), J_-(y^-) \right\} = \frac{1}{2}\, \del^{\prime}(x^- - y^-), }\\
 \\
\dsp{ \left\{ T_{--}(x^-), T_{--}(y^-) \right\} = T_{--}(x^-)\, \del^{\prime}(x^- - y^-)
 - T_{--}(y^-)\, \del^{\prime}(y - x). }
\ea
\label{3.23}
\ee
showing that the central charge is the same as that of the current of
a free conformal scalar theory.

The result (\ref{3.20}) is of considerable interest, as it describes an
infinite-dimensional classical equivalent of a non-commutative geometry
of the complex plane (a non-commutative bundle), expressed by the
non-vanishing bracket of $z(x)$ and $\bz(y)$. For comparison we
recall the strong-coupling limit of a charged particle in a magnetic field,
which is represented by the brackets
\be
\left\{ x^i, x^j \right\} = \lh F^{-1} \rh^{ij}, \hs{2} F_{ij} = \eps_{ijk} B_k,
\label{3.20.1}
\ee
(for unit charge-to-mass ratio). In this standard example of a physical
application of non-commutative geometry the bracket is proportional to the
inverse of the magnetic field, whilst in our case it is proportional to the
inverse of the current component $J_-(x^-)$.

\section{Left-moving fields and currents \label{s.4}}

With the aid of the brackets (\ref{3.18}) we can compute the light-cone
time dependence of the fields using eq.\ (\ref{3.11}). Explicitly we find
\be
\ba{lll}
\der_+ \thg(x) & = & \dsp{ \left\{ \thg(x), S \right\} }\\
 & & \\
 & = & \dsp{ - \frac{1}{J_-(x)} \lh z(x)\, \frac{\del S}{\del z(x)}
 + \bz(x)\, \frac{\del S}{\del \bz(x)} \rh
 - \frac{1}{2}\, \int_{-\infty}^{x^-} dy^-\, \frac{\del S}{\del \thg(x^+, y^-)}, }\\
 & & \\
\der_+ z(x) & = & \dsp{ \left\{ z(x), S \right\}
 = \frac{-i }{4J_-} \lh i z \frac{\del S}{\del \thg(x)} + \frac{\del S}{\del \bz(x)} \rh, }\\
 & & \\
\der_+ \bz(x) & = & \dsp{ \left\{ \bz(x), S \right\}
= \frac{i }{4J_-} \lh -i \bz \frac{\del S}{\del \thg(x)} + \frac{\del S}{\del z(x)} \rh. }
\ea
\label{4.1}
\ee
Therefore in the classical theory, where the action takes an extremal value
and all functional derivatives of the action vanish: $\del S/\del \fg_i = 0$,
the equations of motion take the form
\be
J_- \der_+ z = J_- \der_+ \bz = 0, \hs{2} \der_- \der_+ \thg = 0.
\label{4.2}
\ee
It follows, that $\thg(x)$ is an ordinary 2-dimensional conformal scalar field,
whereas the complex fields $(\bz, z)$ are constrained to have right-moving
components only whenever the current does not vanish identically. Indeed,
taking $J_- \neq 0$ we  find
\be
\der_+ z = \der_+ \bz = 0,  \hs{2} J_+ = \der_+ \thg,
\label{4.3}
\ee
with $\der_- J_+ = 0$. It follows, that after compactification on a circle the
real scalar field has a standard mode expansion
\be
\thg(x^+, x^-) = \thg_0 + \frac{p_0}{4\pi} \lh x^+ + x^- \rh +
 \sum_{n \neq 0} \lh \thg_{+,n} e^{2\pi i n x^+} + \thg_{-, n} e^{-2 \pi i n x^-} \rh,
\label{4.4}
\ee
with the brackets
\be
\left\{ \thg_0, p_0 \right\} = 1, \hs{1}
\left\{ \thg_{+,m}, \thg_{+,n} \right\} = \frac{i}{8 \pi^2 m}\, \del(m+n), \hs{1}
\left\{ \thg_{-,m}, \thg_{-,n} \right\} = \frac{-i}{8 \pi^2 m}\, \del(m+n).
\label{4.5}
\ee
In contrast, the mode expansion of the complex scalars is of the form
\be
z(x^-) = \sum_n z_n e^{2\pi i n x^-}, \hs{2}
\bz(x^-) = \sum_n \bz_n e^{- 2\pi i n x^-},
\label{4.6}
\ee
and these fields do not contribute to $J_+$. It is straightforward to check,
that the brackets of $\thg$ and its derivatives lead to the results
\be
\left\{ \thg(x^+), J_+(y^+) \right\} = \frac{1}{2}\, \del (x^+ - y^+), \hs{2}
\left\{ \thg(x^+), T_{++}(y^+) \right\} = \der_+ \thg\, \del(x^+ - y^+).
\label{4.7}
\ee
In fact, in view of eqs.\ (\ref{4.3}) and
\be
\left\{ z(x^+), J_+(y^+) \right\} = 0, \hs{2} \left\{ \bz(x^+), J_+(y^+) \right\} = 0,
\label{4.7.1}
\ee
we trivially have
\be
\left\{ \fg_i(x^+), T_{++}(y^+) \right\} = \der_+ \fg_i\, \del(x^+ - y^+)
\label{4.8}
\ee
for all elementary fields. Finally, the left-moving currents have the same
central charge as the right-moving ones:
\be
\left\{ J_+(x^+), J_+(y^+) \right\} = \frac{1}{2}\, \del^{\prime}(x^+ - y^+).
\label{4.9}
\ee

\section{Reparametrizations of the target space \label{s.5}}

The equations of motion (\ref{4.3}) and the conservation of the currents
$J_-$ imply, as in sect.\ \ref{s.1}, that there is an infinite set of currents
$J_-[G] = 2 G(\bz, z) J_-$ satisfying the conservation laws
\be
\der_+ J_-[G] = 0.
\label{5.1}
\ee
However, due to the presence of the Wess-Zumino term (\ref{2.5})
these currents do not represent the Noether currents for the
the reparametrization invariance (\ref{1.14}), and do not satisfy simple
bracket relations. Indeed, to obtain the Noether currents in the presence
of the Wess-Zumino term we have to add specific improvement terms to
the currents; if we take $G(\bz, z)$ to be given by a double power series
\be
G(\bz, z) = \sum_{m,n \geq 0}\, g_{mn}\, \bz^m z^n,  \hs{2}
       \bar{g}_{nm} = g_{mn},
\label{5.2}
\ee
we find (for $\lb = +1$) the conserved Noether charges
\be
Q[G] = 4 \int dx^-\, \left[ G J_- - i \sum_{m,n \geq 0}
 \frac{g_{mn}}{(m+1)(n+1)} \lh \bz^{m+1} \lrder_- z^{n+1} \rh \right].
\label{5.3}
\ee
The brackets of these charges indeed satisfy
\be
\ba{l}
\left\{ \thg, Q[G] \right\} = 2 G - z G_{,z} - \bz G_{,\bz}, \\
 \\
\left\{ z, Q[G] \right\} = - i G_{,\bz}, \hs{2} \left\{ \bz, Q[G] \right\} =  i G_{,z}.
\ea
\label{5.4}
\ee
It is then elementary to establish that the bracket algebra of the charges is
closed and of the form (\ref{1.14.1}):
\be
\left\{ Q[G_1], Q[G_2] \right\} = Q[G_3],
\label{5.5}
\ee
with
\be
G_3 = - i \lh \dd{G_1}{z} \dd{G_2}{\bz} - \dd{G_1}{\bz} \dd{G_2}{z} \rh.
\label{5.6}
\ee
In fact eq.\ (\ref{1.14.1}) is equivalent to the Jacobi identity
\be
\left\{ \left\{ A(\fg), Q[G_1] \right\}, Q[G_2] \right\} -
 \left\{ \left\{ A(\fg), Q[G_2] \right\}, Q[G_1] \right\} =
 \left\{ A(\fg), \left\{ Q[G_1], Q[G_2] \right\} \right\},
\label{5.7}
\ee
for arbitrary polynomials $A(\fg)$ of the fields $\fg^i = (\thg, z, \bz)$.

Of particular interest is the existence of two finite-dimensional subalgebras.
The first one is the Heisenberg algebra $e(2)$, generated by the real polynomials
\be
G_e = \frac{1}{2}, \hs{1} G_+ = \frac{1}{2} \lh z + \bz \rh, \hs{1}
G_- = \frac{1}{2i} \lh z - \bz \rh.
\label{5.10}
\ee
They generate the algebra of charges $Q_i  \equiv Q[G_i]$:
\be
\left\{ Q_+, Q_- \right\} = Q_e, \hs{2} \left\{ Q_e, Q_{\pm} \right] = 0.
\label{5.11}
\ee
The second finite-dimensional subalgebra is an $su(1,1)$ algebra generated
by the real quadratic  polynomials
\be
G_1 = \frac{1}{4} \lh z^2 + \bz^2 \rh, \hs{1} G_2 = \frac{1}{4i} \lh z^2 - \bz^2 \rh,
 \hs{1} G_0 = \frac{1}{2}\, \bz z.
\label{5.8}
\ee
Using a similar notation for $Q[G_i]$ we have
\be
\left\{ Q_1, Q_2 \right\} = Q_0, \hs{1}
\left\{ Q_2, Q_0 \right\} = - Q_1, \hs{1}
\left\{ Q_0, Q_1 \right\} = - Q_2.
\label{5.9}
\ee
As mentioned, base-space representations of the algebra SDiff$(M_2)$
of area-preserv\-ing diffeomorphisms of 2--dimensional manifolds $M_2$
have been studied before \ct{arnold,zachos}, in particular in the context 
of membrane theory in ref.\ \ct{hoppe, dewit-nicolai}.

\section{The 2-dimensional conformal quantum fluid \label{s.6}}

In the previous sections we have  established the complete canonical structure
of the 2-dimen\-sional conformal fluid model. In particular, we have determined
the value of the central charge of the current algebra (\ref{3.23}), (\ref{4.9}),
which is identical with that of a free conformal scalar field. If we take the
current algebra as the starting point of the quantization, we can simply
follow the standard Sugawara construction \ct{goddard-olive}, and obtain the
commutator algebra of the energy-momentum tensor. It takes the form of a
Virasoro algebra with central charge $c = 1$.

The only technical detail needed is the operator-ordering prescription. We
briefly recall the procedure, taking the left-moving sector to be specific.
After decomposition of the current into its Fourier modes
\be
J_+ = \sum_n\, J_{+, n} e^{2\pi i n x^+},
\label{6.1}
\ee
quantization of the current algebra (\ref{4.9}) leads to the results
\be
\left[ J_+(x^+), J_+(y^+) \right] = \frac{i}{2}\, \del^{\prime}(x^+ - y^+) \hs{1}
\Leftrightarrow \hs{2} \left[ J_{+,m}, J_{+,n} \right] = - \frac{m}{2}\, \del(m+n).
\label{6.2}
\ee
Clearly, it is essential for operator ordering to separate the positive and
negative frequency components. In particular,  in the quantum theory
the energy momentum tensor is taken to be normal ordered with respect
to this decomposition:
\be
T_{++}(x^-) =\; : J_+(x^-) J_+(x^-) :\, =
 \sum_{n,m} : J_{+, m} J_{+,n} : e^{2\pi i (m+n) x^+}
 = \sum_n T_{+,n} e^{2\pi in x^+},
\label{6.3}
\ee
where
\be
: J_{+,m} J_{+,n} :\, = \left\{ \ba{ll} J_{+,m} J_{+,n}, & m \leq n; \\
                                                J_{+,n} J_{+,m}, & m > n. \ea \rd,
\label{6.4}
\ee
The components of the energy momentum then satisfy the Virasoro algebra
with a central charge determined by the central charge of the current algebra
(\ref{3.23}):
\be
\left[ T_{+,m}, T_{+,n} \right] = (m - n)\, T_{+, m+n} + \frac{1}{12}\, (m^3 -m)\,
 \del(m+n),
\label{6.5}
\ee
It remains to establish the current algebra result  (\ref{6.2}) in the quantum
fluid model. For the left-moving sector this is trivial, as the current simply
is that of a free scalar field
\be
J_+ = \frac{p_0}{4\pi} + \sum_{m \neq 0} 2 \pi i m\, \thg_{+,m}\, e^{2\pi i m x^+},
\label{6.6}
\ee
where the scalar field satisfies the commutation relations
\be
\left[ \thg_{+,m}, \thg_{+,n} \right] = \frac{-1}{8\pi^2 m}\, \del(m+n).
\label{6.7}
\ee
This leads directly to the desired result (\ref{6.2}).

However, in the sector of right-moving fields the situation is complicated by
the fact that the non-linear brackets (\ref{3.20}) between the complex fields
$\bz$ and $z$, and between these fields and the real scalar $\thg$, prevent
a decomposition into Fourier modes. It also poses a problem of operator
ordering for the expressions on the right-hand side of the brackets.
For example, proceeding in co-ordinate space and choosing a particular
ordering to postulate the commutation rules
\be
\ba{ll}
\dsp{ \left[ \thg(x^-), \thg(y^-) \right] = - \frac{i}{2}\, \eps(x^- - y^-), }&
\dsp{ \left[ z(x^-), \bz(y^-) \right] = \frac{1}{4}\, J^{-1}_-\, \del(x^- - y^-), }\\
 & \\
\dsp{ \left[ z(x^-), \thg(y^-) \right] = \frac{i}{4}\, J^{-1} z\, \del(x^- - y^-), }&
\dsp{ \left[ \bz(x^-), \thg(y^-) \right] = \frac{i}{4}\, \bz J^{-1}\, \del(x^- - y^-), }
\ea
\label{6.7.1}
\ee
whilst defining the ordered current by
\be
J_- = \der_- \thg + i \bz \der_- z - i \der_- \bz\, z,
\label{6.7.0}
\ee
a straighforward calculation shows that
\be
\left[ J_-(x^-), J_-(y^-) \right] = \frac{i}{2}\, \del^{\prime}(x^- - y^-).
\label{6.7.2}
\ee
More precisely, introducing smooth bounded test functions $f(x)$ and $g(x)$
we have established that
\be
J_f = \int dx^-\, f(x^-) J_-(x^-) \hs{1} \Rightarrow \hs{1}
\left[ J_f, J_g \right] = \frac{i}{2}\, \int dx^-\, f \der_- g.
\label{6.7.3}
\ee
This implies that the central charges of the current algebra and the algebra
of the energy-momentum tensor indeed have the same values as in the
left-moving sector, i.e.\ we have a free field theory with $c = 1$.

Unfortunately, the proposed set of commutation relations (\ref{6.7.1}) seems
to be inconsistent; indeed, a check of the Jacobi identities gives
\be
\left[ \bz(x), \left[ z(y), \thg(u) \right] \right] +
\left[ z(y), \left[ \thg(u), \bz(x) \right] \right] +
\left[ \thg(u), \left[ \bz(x), z(y) \right] \right]  = i q(x)\, \del(x-y) \del(u-y),
\label{6.9}
\ee
where $q$ is a divergent operator defined by
\be
q = \frac{1}{4}\, \mbox{Im}\, \lh J^{-1} \der \bz J^{-2} z - J^{-1} \der\bz J^{-1}
 z J^{-1} \rh = \frac{1}{8}\, \del(0)\, J^{-1} \der \bz J^{-3} \der z J^{-1}.
\label{6.10}
\ee
This may well be due to our choice of operator ordering, both in the
commutators and in the definition of the current $J_-$. In appendix
\ref{a.2} we consider a more general approach, allowing a larger class
of operator orderings. In all cases we find the same value of the
central charge in the current algebra (\ref{6.7.2}). However, for none
of the orderings we considered are the Jacobi identities manifestly
satisfied. We can therefore not exclude that the quantum theory is
anomalous in the sense of the Groenewold-Van Hove theorem
\ct{groenewold,vanhove,gotay}.

\section{Discussion \label{s.7}}

In this paper we have presented a model of conformal fluids coupled to
general relativity, formulated in $n$ space-time dimensions; we have
focussed on the case $n = 2$, where the well-known techniques of
conformal field theory can act as a guide to the construction of a full
quantum theory. A point of interest in these models is the existence of
an infinite set of conserved charges generating area-preserving
diffeomorphisms in the target space of the complex scalar potentials
$(\bz, z)$. This algebra is preserved by the Wess-Zumino term in the
2-dimensional theory, and a non-linear canonical realization of the
diffeomorphism algebra has been constructed in terms of light-cone
Poisson brackets as a first step towards a non-commutative geometry
of a complex bundle over the circle.

We have carried out the construction for the simplest case, in which
the scalar fields $(\bz, z)$ take their values in the complex plane.
However, it is easy to construct models where they take values in
other K\"{a}hler manifolds. This is achieved by replacing the
current one-form (\ref{1.15}) by the generalized expression
\be
J = d \thg + i K_{,z}\, dz - i K_{,\bz\,} d \bz,
\label{7.1}
\ee
with $K(\bz,z)$ the (real) K\"{a}hler potential of the manifold \ct{nyawetal}.
Such different potential manifolds may for example lead to different central
charges in the diffeomorphism algebra \ct{dewit-nicolai,zachos2}.

Finally, we have discussed aspects of the quantization of the models.
We have found that the non-linearity of the Poisson brackets (\ref{3.20})
complicates a direct translation to commutators in the full theory,
incorporating the complex scalars and their diffeomorphism invariance.
On the level of currents and energy-momentum components this is not
an issue, as the quantization of the current algebra immediately
implies the Virasoro algebra by the Sugawara construction. In addition,
one can also quantize the theory after elimination of the complex scalars
as a simple free theory of a single real conformal scalar $\thg$ with
standard values of the central charges. However,  to construct a quantum
realization of the diffeomorphism algebra (\ref{5.5}), (\ref{5.6}), and hence
a complete non-commutative geometry of operators $z(x)$ and $\bz(y)$
taking values in the complex plane or some other K\"{a}hler manifold, the
full set of classical brackets must be mapped consistently to operator
commutators, a problem we have not solved. If a solution exists, it will be
of interest to compare such a construction with the Moyal bracket approach
as explained in \ct{zachos2}.
\vs{3}

\nit
{\bf Acknowledgement} \\
J.W.v.H.\ wishes to thank the School of Mathematics and Physics
of the University of Tasmania in Hobart, where a large part of this work
was performed, for its hospitality and support. This work was partially
supported by the Australian Research Council, project DP-0208808.

\np
\appendix
\section{Components of the symplectic form \label{a.1}}

In this appendix we collect the explicit expressions for the components
of the symplectic tensor $F_{ij}(x^+; x^-, y^-)$ in eq.\ (\ref{3.9}). As
all fields are always evaluated at equal light-cone time $x^+$, in the
following we only make explicit reference to the light-cone space
co-ordinate, i.e.\ $(x, y, ...) = (x^-, y^-, ...)$. We then have
\be
\ba{lll}
F_{\thg \thg}(x,y)  =  - F_{\thg \thg}(y,x)  & = & - 2\, \del^{\prime}(x-y), \\
 & &  \\
F_{\thg z}(x,y) = - F_{z \thg}(y,x) & = & - 2i \bz(y)\, \del^{\prime}(x-y)  
 + 2 i (1 - \lb) \der_- \bz\, \del(x-y), \\
 & & \\
F_{\thg \bz}(x,y) = - F_{\bz \thg}(y,x) & = & 2i z(y)\, \del^{\prime}(x-y)
 - 2i (1 - \lb) \der_- z\, \del(x - y), \\
 & & \\
F_{z z}(x,y) =  - F_{z z}(y,x)  & = & \lh \bz^2(x) + \bz^2(y) \rh 
 \del^{\prime}(x - y) = 2 \bz(x) \bz(y)\, \del^{\prime}(x-y), \\
 & & \\
F_{z \bz}(x,y) = - F_{\bz z}(y,x) & = & - \lh \bz(x) z(x) + \bz(y) z(y) \rh 
 \del^{\prime}(x - y) \\
 & & \\
 & &  +\, i \lh - 2 J_- - i \bz \lrder_- z - 2 \lb \der_- \thg \rh \del(x - y) \\
 & & \\
 & = & -2 \bz(x) z(y)\, \del^{\prime}(x - y) \\
 & & \\
 & & -\, 2i \lh  (1 + \lb) J_- + (1 - \lb) i \bz \lrder_- z \rh  \del(x - y), \\
 & & \\
F_{\bz\bz}(x,y) = - F_{\bz\bz}(y,z) & = & \lh z^2(x) + z^2(y) \rh 
 \del^{\prime}(x - y) = 2 z(x) z(y)\, \del^{\prime}(x - y).
\ea
\label{a1.1}
\ee

\section{The current and its commutators  \label{a.2}}

In this appendix we derive some results on commutators of the complex
scalar fields $\bz$, $z$ and the right-moving current $J_-$, relevant to a 
discussion of the quantization of the brackets (\ref{3.20}), (\ref{3.21}). 
We start from a minimal Ansatz, by postulating the operator commutators
\be 
\ba{l}
\dsp{ \left[ z(x), \bz(y) \right] = \frac{1}{4}\, J^{-1}(x)\, \del(x - y), }\\
 \\
\dsp{ \left[ z(x), J(y) \right] = \frac{i}{2} \lh (1 - \kg)\, \der z\, J^{-1}
 + \kg J^{-1} \der z \rh \del(x - y), }\\
 \\
\dsp{ \left[ \bz(x), J(y) \right] = \frac{i}{2} \lh (1 - \kg)\, J^{-1} \der \bz
 + \kg \der \bz J^{-1}\rh \del(x - y), }
\ea
\label{a2.1}
\ee 
Here all arguments refer to right-moving light-cone co-ordinates $(x^-, y^-)$,
and $J(x)$ denotes the right-moving current. The parameter $\kg$ is
introduced to take into account some of the ordering ambiguities in the
conversion of the classical brackets to quantum commutators. We
postulate no other commutation relations; that is, we choose to eliminate
the real scalar field $\thg$ in favor of the current $J$ as an elementary
operator in the theory. Of course, we then also have to know the commutator
of two current operators. However, instead of postulating this commutator,
we will try to derive it from consistency of the postulated commutators
(\ref{a2.1}). We proceed in several steps.

First, we observe that the commutation relations (\ref{a2.1}) satisfy the
Jacobi identity
\be
\left[ z(x),\left[ z(y), \bz(u) \right] \right] +
 \left[ z(y),\left[\bz(u), z(x) \right] \right] +
 \left[ \bz(u),\left[ z(x), z(y) \right] \right] = 0.
\label{a2.2}
\ee
Next we consider the Jacobi identity
\be
\left[ z(x),\left[ \bz(y), J(u) \right] \right] +
 \left[ \bz(y),\left[ J(u), z(x) \right] \right] +
 \left[ J(u), \left[ z(x), \bz(y) \right] \right]= 0.
\label{a2.3}
\ee
By substitution of the Ansatz (\ref{a2.1}) we find that
this identity holds if and only if
\be
\ba{rcl}
\left[ J(x), J(y) \right] & = & \dsp{ \frac{i}{2}\, \del^{\prime} (x - y) + i C(x)\, \del(x-y),}\\
 & & \\
C(x) & = & \dsp{ \mbox{Im} \left[
  (1 - \kg)^2 \der z J^{-2} \der \bz J + \kg^2  J \der \bz J^{-2} \der z \rd }\\
 & & \\
 & & \dsp{ \ld ~~~+\, \kg (1 - \kg) \lh J \der \bz J^{-1} \der z J^{-1}
         + J^{-1} \der z J^{-1} \der \bz J \rh \right] \del (x - y). }
\ea
\label{a2.4}
\ee
This result has several interesting properties. First, the central charge
term proportional to the derivative of the Dirac $\del$-function is a $c$-number
independent of $\kg$; therefore it does not depend on the operator ordering.
Also, it has the same value as in the classical bracket, implying  that in a
properly defined quantum theory the value of this central charge and the central
charge in the Virasoro algebra will be the same as in the classical theory,
i.e.\ $c = 1$. Finally, in the classical theory where all quantities are ordinary
functions rather than operators, the quantity inside the square brackets
defining $C(x)$ is real; hence in the classical theory $C(x) = 0$, as we
found indeed in sect.\ \ref{s.3}.

For the quantum theory with commutation relations (\ref{a2.1}) to exist,
 it is necessary that $C(x) = 0$ as an operator equation. This follows
directly from the identity
\be
\left[ J(x), J(x) \right] = 0,
\label{a2.5}
\ee
or its regularized equivalent
\be 
\left[ J_f, J_f \right] = 0,
\label{a2.6}
\ee
where
\be
J_f = \int dx\, f(x) J(x),
\label{a2.7}
\ee
with $f(x)$ a smooth test function. This identity states that the current is a
well-behaved local operator and requires $C(x)$ to vanish. Therefore if the
commutation relations (\ref{a2.1}) are consistent for some value of $\kg$,
then the classical bracket for the currents immediately carries over to the
quantum commutator and we have
\be
\left[ J(x), J(y) \right] = \frac{i}{2}\, \del^{\prime} (x - y).
\label{a2.8}
\ee
With this result, all the other Jacobi identities turn out to be satisfied as well;
for example we find the non-trivial result that
\be
\left[ J(x), \left[ J(y), z(u) \right] \right] +
  \left[ J(y), \left[ z(u), J(x)\right] \right] +
  \left[ z(u), \left[ J(x), J(y) \right] \right] = 0.
\label{a2.9}
\ee
Then we end up with a consistent quantum theory in which the central
charge of the current commutator has the classical value. However,
it is not at all clear how the consistency requirement $C(x) = 0$
follows from a direct calculation. In sect.\ \ref{s.6} we produced the
result (\ref{a2.8}) for $\kg = 1$, but only at the expense of violating
another, closely related Jacobi identity.

As an aside, we note that one could also take a short cut by observing
from the classical theory, that the current algebra is not changed by
taking $\bz = z = 0$ both in the left- and right-moving sector. In this
sense, the complex scalars are only auxiliary fields; the diffeomorphism
invariance in the complex target space is a manifestiation of this property.
Upon making this simplification, the theory becomes that of a single
real massless scalar field, which can be quantized in the standard way.
However, in this case the diffeomorphism invariance is trivially removed,
and there is no possibility of constructing a quantum version of the
infinite-dimensional algebra (\ref{5.3}), (\ref{5.5}).

\np

\end{document}